\documentclass[preprint,aps,12pt,showpacs,nofootinbib,tightenlines]{revtex4}
\usepackage{amsmath}
\usepackage{amssymb}
\usepackage{epsfig}
\usepackage{graphicx}
\begin{document}
\def\be{\begin{eqnarray}}
\def\en{\end{eqnarray}}
\def\non{\nonumber\\}

\def\ra{\rangle}
\def\la{\langle}
\def\sl{\!\!\!\slash}
\def\prd{{Phys. Rev. D}~}
\def\prl{{ Phys. Rev. Lett.}~}
\def\plb{{ Phys. Lett. B}~}
\def\npb{{ Nucl. Phys. B}~}
\def\epjc{{ Eur. Phys. J. C}~}
\newcommand{\acp}{{\cal A}_{CP}}

\newcommand{\psl}{ P \hspace{-2.4truemm}/ }
\newcommand{\nsl}{ n \hspace{-2.2truemm}/ }
\newcommand{\vsl}{ v \hspace{-2.2truemm}/ }
\newcommand{\epsl}{\epsilon \hspace{-1.8truemm}/\,  }
\title{Study of $f_0(980)$ and $f_0(1500)$ from $ B_s \to f_0(980) K, f_0(1500) K$ Decays}
\author{Zhi-Qing Zhang
\footnote{Electronic address: zhangzhiqing@haut.edu.cn} } 
\affiliation{\it \small  Department of Physics, Henan University of Technology,
Zhengzhou, Henan 450052, P.R.China }
\date{\today}
\begin{abstract}
In this paper, we calculate the branching ratios and CP-violating
asymmetries for $\bar B^0_s \to f_0(980)K, f_0(1500)K$ within Perturbative QCD approach based on $k_T$
factorization. If the mixing angle $\theta$ falls into the range of $25^\circ<\theta<40^\circ$, the branching ratio of
$\bar B^0_s\to f_0(980)K$ is $ 2.0\times 10^{-6}<{\cal B}(\bar B^0_s\to f_0(980)K)<2.6\times 10^{-6}$,
while $\theta$ lies in the range of $140^\circ<\theta<165^\circ$, ${\cal B}(\bar B^0_s\to f_0(980)K)$ is about
$6.5\times 10^{-7}$. As to the decay ${\cal B}(\bar B^0_s\to f_0(1500)K)$, when the mixing scheme $\mid f_0(1500)\rangle=0.84\mid s\bar s
\rangle-0.54\mid n\bar n \rangle$ for $f_0(1500)$ is used, it is difficult to determine which scenario is more preferable
than the other one from the branching ratios for these two scenarios, because they are both close to $1.0\times10^{-6}$. But there exists large difference in the form factor
$F^{\bar B_s^0\to f_0(1500)}$ for two scenarios.
\end{abstract}

\vspace{1cm}

\maketitle


\section{Introduction}\label{intro}

For scalars' mysterious structure, it arose much interest
in both theory and experiment. In order to uncover the inner structures, many
approaches are used to research the $B_{u,d}$ decay modes with a scalar meson in the final states,
such as the generalized factorization approach \cite{GMM}, QCD
factorization approach (QCDF) \cite{CYf0K,CCYscalar,CCYvector}, Perturbative
QCD (PQCD) approach \cite{Chenf0K1,Chenf0K2,wwang,zqzhang1,zqzhang2,ylshen}. But as to $B^0_s$ meson, these decay modes
haven not been well studied by
theory. The role of scalar particles in $B^0_s$ decays should be given much more noticeable, because analyses
of the corresponding decays can also provide a unique insight to the mysterious structure of the scalar mesons.
Here we we will study the branching ratios and CP asymmetries of $\bar B^0_s \to f_0(980)K, f_0(1500)K$ within Perturbative
QCD approach. The fundamental concept of this approach is factorization theorem, which states that
the nonperturbative dynamics pross can be separated from a high-energy QCD process. The remaining part, being
infrared finite, is calculable in perturbation theory. So a full amplitude is expressed as the convolution
of perturbative hard kernels with hadron wave functions. There is a parton momentum fraction $x$ in the former, both
$x$ and $k_T$ in the latter. Because $x$ must be integrated over in the range between $0$ and $1$, the end-point
region with a small $x$ is not avoidable. If there is a singularity developed in a formula, $k_T$ factorization
should be employed \cite{sterman1,sterman2}. Here $k_T$ denotes parton transverse mometa. A wave function, because of
its nonperturbative origin, is not calculable, but process independent. So it can be determined by some means, such
as QCD sum rules and lattice theory or extracted from experimental data. On the experimental side,
some of $B^0_s$ decays involved a
scalar in the final states might be observed in the Large Hadron Collider beauty experiments (LHC-b) \cite{lhc1,lhc2}.
In order to make precision studies of rare decays in the B-meson systems, the LHC-b
detector is designed to exploit the large number of b-hadrons produced. Furthermore, it can reconstruct a B-decay
vertex with very good resolution, which is essential for studying the rapidly oscillating $B_s$ mesons.
So the studies of these decay modes of $B^0_s$
are necessary in the next a few years.

It is organized as follows: In Sect.\ref{proper}, we introduce the input parameters including the decay constants
and light-cone distribution amplitudes.  In Sec.\ref{results}, we
then apply PQCD approach to calculate analytically the branching
ratios and CP asymmetries for our considered decays. The final part
contains our numerical results and discussions.


\section{Input Parameters}\label{proper}
For the  underlying structure of the scalar mesons is still under
controversy, there are two typical schemes for the classification to
them \cite{nato,jaffe}. The scenario I (SI): the nonet mesons below
1 GeV, including $f_0(600), f_0(980), K^*(800)$ and $a_0(980)$, are
usually viewed as the lowest lying $q\bar q$ states, while the nonet
ones near 1.5 GeV, including $f_0(1370), f_0(1500)/f_0(1700),
K^*(1430)$ and $a_0(1450)$, are suggested as the first excited
states. In the scenario II (SII), the nonet mesons near 1.5 GeV are
treated as $q\bar q$ ground states, while the nonet mesons below 1
GeV are exotic states beyond the quark model such as four-quark
bound states. In order to make quantitative predictions, we identify
$f_0(980)$ as a mixture of $s\bar s$ and $n\bar n=(u\bar u+d\bar
d)/\sqrt2$, that is
 \be |f_0(980)\ra = |s\bar s\ra\cos\theta+|n\bar
n\ra\sin\theta, \en where the mixing angle $\theta$ is taken in the
ranges of $25^\circ< \theta <40^\circ$ and
$140^\circ<\theta<165^\circ$ \cite{hycheng}. Certainly, $f_0(1500)$
can be treated as a $q\bar q$ state in both SI and  SII. We
considered that the meson $f_0(1500)$ and $f_0(980)$ have the same
component structure but with different mixing angle.

For the the neutral scalar mesons $f_0(980)$ and $f_0(1500)$
cannot be produced via the vector current, we have $\langle f_0(p)|\bar q_2\gamma_\mu q_1|0\ra=0$.
Taking the mixing into account, the scalar current $\langle f_0(p)|\bar q_2q_1|0\ra=m_S\bar {f_S}$
can be written as:
\be
\langle f_0^n|d\bar
d|0\ra=\langle f_0^n|u\bar u|0\ra=\frac{1}{\sqrt 2}m_{f_0}\tilde
f^n_{f_0},\,\,\,\, \langle f_0^n|s\bar s|0\ra=m_{f_0}\tilde
f^s_{f_0},
\en
where $f_0^{(n,s)}$ represent for the light-cone
distribution amplitudes for $n\bar{n}$ and $s\bar{s}$ components,
respectively. Using the QCD sum rules method, one can find the
scale-dependent scalar decay constants $f_{f_0}^n$ and $f_{f_0}^s$
are very close \cite{CCYscalar}. So we shall assume $\tilde
f_{f_0}^n=\tilde f_{f_0}^s$ and denote them as $\bar f_{f_0}$ in the
following.

The twist-2 and twist-3 light-cone distribution amplitudes (LCDAs)
for different components of $f_0$ are defined by
\be
\langle f_0(p)|\bar q(z)_l q(0)_j|0\rangle
&=&\frac{1}{\sqrt{2N_c}}\int^1_0dxe^{ixp\cdot z}\{p\sl\Phi_{f_0}(x)
+m_{f_0}\Phi^S_{f_0}(x)+m_{f_0}(n\sl_+n\sl_--1)\Phi^{T}_{f_0}(x)\}_{jl},\non
\label{LCDA}
\en
here we assume $f_0^n(p)$ and $f_0^s(p)$ are same and denote
them as $f_0(p)$, $n_+$ and $n_-$ are light-like vectors:
$n_+=(1,0,0_T),n_-=(0,1,0_T)$. The normalization can be related to
the decay constants
\be \int^1_0 dx\Phi_{f_0}(x)=\int^1_0
dx\Phi^{T}_{f_0}(x)=0,\,\,\,\,\,\,\,\int^1_0
dx\Phi^{S}_{f_0}(x)=\frac{\bar f_{f_0}}{2\sqrt{2N_c}}.
\en

The wave function for $K$ meson is
given as \be \Phi_{K}(P,x,\zeta)\equiv
\frac{1}{\sqrt{2N_C}}\gamma_5 \left [ \psl \Phi_{K}^{A}(x)+m_0^{K}
\Phi_{K}^{P}(x)+\zeta m_0^{K} (\vsl \nsl - v\cdot
n)\Phi_{K}^{T}(x)\right ], \en
where $P$ and $x$ are the momentum and the momentum fraction of
$K$ meson, respectively. The parameter $\zeta$ is either $+1$ or $-1$ depending on the
assignment of the momentum fraction $x$.

In general, the $B_s$ meson is treated as heavy-light system and its Lorentz structure
can be written as \cite{grozin,kawa}
\be
\Phi_{B_s}=\frac{1}{\sqrt{2N_c}}(\psl_{B_s}+M_{B_s})\gamma_5\phi_{B_s}(k_1).
\en
For the contribution of $\bar \phi_{B_s}$ is numerically small \cite{caidianlv} and has been neglected.
\section{Theoretical Framework and perturbative calculations} \label{results}
Under the two-quark model for the scalar mesons supposition, we
would like to use PQCD approach to study $\bar B^0_s \to f_0(980)K, f_0(1500)K$ decays.
In this approach, the decay amplitude is
separated into soft, hard, and harder dynamics characterized by
different energy scales $(t, m_{B_s}, M_W)$. It is conceptually
written as the convolution,
\be
{\cal A}(\bar B^0_s \to f_0K)\sim \int\!\!
d^4k_1 d^4k_2 d^4k_3\ \mathrm{Tr} \left [ C(t) \Phi_{B_s}(k_1)
\Phi_{f_0}(k_2) \Phi_{K}(k_3) H(k_1,k_2,k_3, t) \right ],
\label{eq:con1}
\en
where $k_i$'s are momenta of anti-quarks included in each meson, and $\mathrm{Tr}$ denotes the trace over
Dirac and color indices. $C(t)$ is the Wilson coefficient which
results from the radiative corrections at a short distance. The function
$H(k_1,k_2,k_3,t)$ describes the four quark operator and the
spectator quark connected by a hard gluon whose $q^2$ is in the order
of $\bar{\Lambda} M_{B_s}$, and includes the $\mathcal{O}(\sqrt{\bar{\Lambda} M_{B_s}})$ hard dynamics.
Therefore, this hard part $H$ can be perturbatively calculated.

Since the b quark is rather heavy, we consider the $\bar B^0_s$ meson at rest
for simplicity. It is convenient to use light-cone coordinate $(p^+,
p^-, {\bf p}_T)$ to describe the meson's momenta,
\be
p^\pm =\frac{1}{\sqrt{2}} (p^0 \pm p^3), \quad {\rm and} \quad {\bf p}_T =
(p^1, p^2).
\en
Using these coordinates the $\bar B^0_s$ meson and the two
final state meson momenta can be written as
\be P_{B_s} =
\frac{M_{B_s}}{\sqrt{2}} (1,1,{\bf 0}_T), \quad P_{2} =
\frac{M_{B_s}}{\sqrt{2}}(1,0,{\bf 0}_T), \quad P_{3} =
\frac{M_{B_s}}{\sqrt{2}} (0,1,{\bf 0}_T), \en respectively. The
meson masses have been neglected. Putting the anti-quark momenta in
$\bar B^0_s$, $f_0$ and $K$ mesons as $k_1$, $k_2$, and $k_3$, respectively, we
can choose
\be
k_1 = (x_1 P_1^+,0,{\bf k}_{1T}), \quad k_2 = (x_2
P_2^+,0,{\bf k}_{2T}), \quad k_3 = (0, x_3 P_3^-,{\bf k}_{3T}).
\en
For our considered decay channels, the integration over $k_1^-$,
$k_2^-$, and $k_3^+$ in equation (\ref{eq:con1}) will lead to
\be {\cal
A}(\bar B^0_s \to f_0K) &\sim &\int\!\! d x_1 d x_2 d x_3 b_1 d b_1 b_2 d
b_2 b_3 d b_3 \non && \cdot \mathrm{Tr} \left [ C(t)
\Phi_{B_s}(x_1,b_1) \Phi_{f_0}(x_2,b_2) \Phi_{K}(x_3, b_3) H(x_i,
b_i, t) S_t(x_i)\, e^{-S(t)} \right ], \quad \label{eq:a2} \en where
$b_i$ is the conjugate space coordinate of $k_{iT}$, and $t$ is the
largest energy scale in function $H(x_i,b_i,t)$. The large
double logarithms ($\ln^2 x_i$) on the longitudinal direction are
summed by the threshold resummation \cite{li02}, and they lead to
$S_t(x_i)$, which smears the end-point singularities on $x_i$. The
last term $e^{-S(t)}$ is the Sudakov form factor which suppresses
the soft dynamics effectively \cite{soft}. Thus it makes the
perturbative calculation of the hard part $H$ applicable at
intermediate scale, i.e., $M_{B_s}$ scale.

We will calculate analytically the function $H(x_i,b_i,t)$ for $\bar B^0_s
\to f_0K$ decays in the leading-order and give the convoluted
amplitudes. For our considered decays, the related weak effective
Hamiltonian $H_{eff}$ can be written as \cite{buras96} \be
\label{eq:heff} {\cal H}_{eff} = \frac{G_{F}} {\sqrt{2}} \,
\sum_{q=u,c}V_{qb} V_{qd}^*\left[ \left (C_1(\mu) O_1^q(\mu) +
C_2(\mu) O_2^q(\mu) \right) \sum_{i=3}^{10} C_{i}(\mu) \,O_i(\mu)
\right] \; , \en with the Fermi constant $G_{F}=1.166 39\times
10^{-5} GeV^{-2}$, and the CKM matrix elements V. We specify below
the operators in ${\cal H}_{eff}$ for $b \to d$ transition \be
\begin{array}{llllll}
O_1^{u} & = &  \bar d_\alpha\gamma^\mu L u_\beta\cdot \bar
u_\beta\gamma_\mu L b_\alpha\ , &O_2^{u} & = &\bar
d_\alpha\gamma^\mu L u_\alpha\cdot \bar
u_\beta\gamma_\mu L b_\beta\ , \\
O_3 & = & \bar d_\alpha\gamma^\mu L b_\alpha\cdot \sum_{q'}\bar
 q_\beta'\gamma_\mu L q_\beta'\ ,   &
O_4 & = & \bar d_\alpha\gamma^\mu L b_\beta\cdot \sum_{q'}\bar
q_\beta'\gamma_\mu L q_\alpha'\ , \\
O_5 & = & \bar d_\alpha\gamma^\mu L b_\alpha\cdot \sum_{q'}\bar
q_\beta'\gamma_\mu R q_\beta'\ ,   & O_6 & = & \bar
d_\alpha\gamma^\mu L b_\beta\cdot \sum_{q'}\bar
q_\beta'\gamma_\mu R q_\alpha'\ , \\
O_7 & = & \frac{3}{2}\bar d_\alpha\gamma^\mu L b_\alpha\cdot
\sum_{q'}e_{q'}\bar q_\beta'\gamma_\mu R q_\beta'\ ,   & O_8 & = &
\frac{3}{2}\bar d_\alpha\gamma^\mu L b_\beta\cdot
\sum_{q'}e_{q'}\bar q_\beta'\gamma_\mu R q_\alpha'\ , \\
O_9 & = & \frac{3}{2}\bar d_\alpha\gamma^\mu L b_\alpha\cdot
\sum_{q'}e_{q'}\bar q_\beta'\gamma_\mu L q_\beta'\ ,   & O_{10} & =
& \frac{3}{2}\bar d_\alpha\gamma^\mu L b_\beta\cdot
\sum_{q'}e_{q'}\bar q_\beta'\gamma_\mu L q_\alpha'\ ,
\label{eq:operators} \end{array} \en where $\alpha$ and $\beta$ are
the $SU(3)$ color indices; $L$ and $R$ are the left- and
right-handed projection operators with $L=(1 - \gamma_5)$, $R= (1 +
\gamma_5)$. The sum over $q'$ runs over the quark fields that are
active at the scale $\mu=O(m_{B_s})$, i.e.,
$(q'\epsilon\{u,d,s,c,b\})$.

\begin{figure}[t,b]
\vspace{-4cm} \centerline{\epsfxsize=20 cm \epsffile{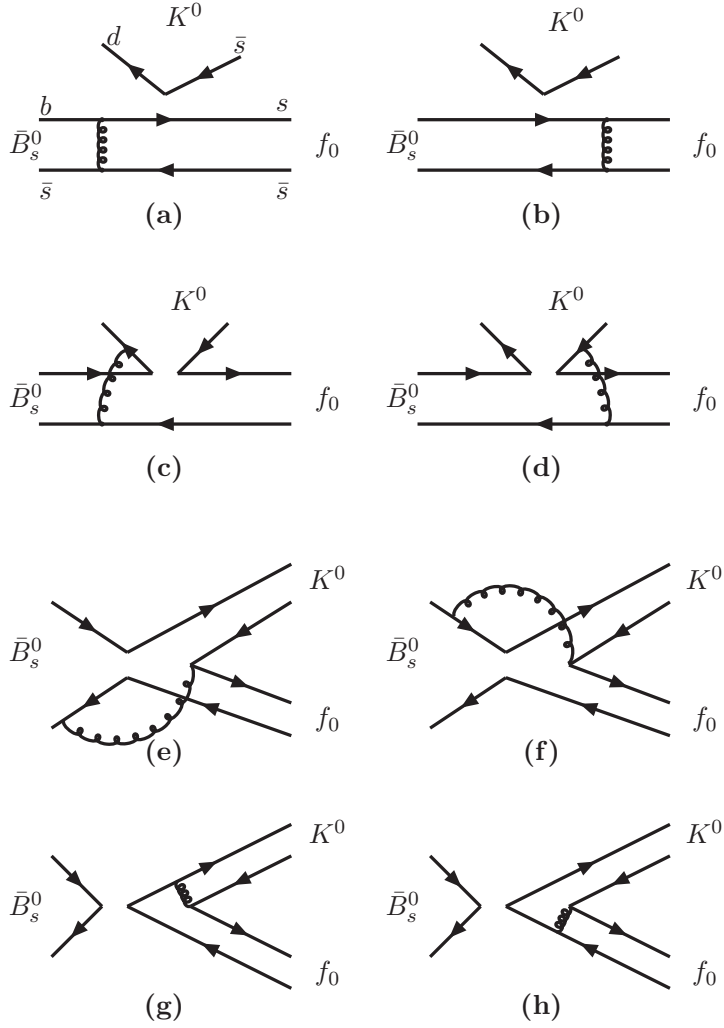}}
\vspace{-10.5cm} \caption{ Diagrams contributing to the $\bar B^0_s\to
f_0K^0$ decays .}
 \label{Figure1}
\end{figure}

There are eight type diagrams contributing to the $\bar B_s \to
f_0K$ decays are illustrated in figure \ref{Figure1}. For the
factorizable emission diagrams (a) and (b), Operators $O_{1,2,3,4,9,10}$ are
$(V-A)(V-A)$ currents, and the operators $O_{5,6,7,8}$ have a
structure of $(V-A)(V+A)$, the sum of the their amplitudes are
written as $F_{ef_0}$ and $F_{ef_0}^{P1}$. In some other cases, we need to do Fierz transformation
for the $(V-A)(V+A)$ operators and get $(S-P)(S+P)$ ones which hold right flavor and color structure
for factorization work. The contribution from operator type $(S-P)(S+P)$ is written as $F_{ef_0}^{P2}$; Similarly,
for the facorizable annihilation diagrams (g) and (h), The contributions from $(V-A)(V-A), (V-A)(V+A), (S-P)(S+P)$ these
three kinds of operators are $F_{af_0}, F_{af_0}^{P1}$ and $F_{af_0}^{P2}$, respectively. For the nonfactorizable emission
(annihilation) diagrams (c) and (d) ((e) and (f)), these three kinds of contributions can be written as $M_{e(a)f_0},
M_{e(a)f_0}^{P1}, M_{e(a)f_0}^{P2}$, respectively. Since these amplitudes are similar
to those for the decays $B \to f_0(980)K(\pi,\eta^{(\prime)})$ \cite{wwang,zqzhang1} or $B \to a_0(980)K$ \cite{ylshen}, we just need to replace
some corresponding wave functions and parameters. It is the same with the amplitudes for the $f_0$ and $K^0$ exchanging diagrams.

Combining the contributions from different diagrams, the total decay
amplitudes for these decays can be written as
\be
{\cal M}
(\bar B^0_s\to f_0K) &=&{\cal M}_{s\bar{s}}\times\cos\theta+\frac{1}{\sqrt{2}}{\cal M}_{n\bar n}\sin\theta,\label{snamp}
\en
with
\be {\cal M}_{s\bar{s}}&=&
-V_{tb}V^*_{td}\left[(F_{ef_0}+F_{af_0})(a_4-\frac{1}{2}a_{10})+(F^{P2}_{ef_0}+F^{P2}_{af_0})(a_6-\frac{1}{2}a_8)\right.\nonumber\\
&&\left.
+(M_{ef_0}+M_{af_0})(C_3-\frac{1}{2}C_9)+(M^{P1}_{ef_0}+M^{P1}_{af_0})
(C_5-\frac{1}{2}C_7)\right.
\nonumber\\
&&\left.
+M_{eK}(C_4-\frac{1}{2}C_{10})+M^{P2}_{eK}(C_6-\frac{1}{2}C_{8})
\right]
, \label{ssbar}\en
\be {\cal M}_{n\bar{n}}&=&
V_{ub}V^*_{ud}M_{eK}C_2-V_{tb}V^*_{td}\left[(F^{P2}_{eK}+F^{P2}_{aK})(a_6-\frac{1}{2}a_8)
+(M^{P1}_{eK}+M^{P1}_{aK})(C_5-\frac{1}{2}C_7)\right.\nonumber\\
&&\left.
+M_{eK}(C_3+2C_4+\frac{1}{2}(C_{10}-C_9))+M^{P2}_{eK}(2C_6+\frac{1}{2}C_8)+M_{aK}(C_3-\frac{1}{2}C_9)
\right.\nonumber\\
&&\left.+F_{aK}(a_4-\frac{1}{2}a_{10})
\right]
, \label{nnbar}\en
where the combinations of the Wilson coefficients are defined as usual
\cite{AKL,keta}
\begin{eqnarray}
a_1= C_2+C_1/3, & a_3= C_3+C_4/3,~a_5= C_5+C_6/3,~a_7=
C_7+C_8/3,~a_9= C_9+C_{10}/3,\quad\quad\\
a_2= C_1+C_2/3, & a_4= C_4+C_3/3,~a_6= C_6+C_5/3,~a_8=
C_8+C_7/3,~a_{10}= C_{10}+C_{9}/3.\quad\quad
\end{eqnarray}

\section{Numerical results and discussions}
The twist-2 LCDA can be expanded in the Gegenbauer polynomials \be
\Phi_{f_0}(x,\mu)&=&\frac{1}{2\sqrt{2N_c}}\bar
f_{f_0}(\mu)6x(1-x)\sum_{m=1}^\infty B_m(\mu)C^{3/2}_m(2x-1),
\en
where $B_m(\mu)$ and $C^{3/2}_m(x)$ are the Gegenbauer moments and Gegenbauer polynomials, respectively. The values for Gegenbauer moments
and the decay constants are taken (at scale $\mu=1 \mbox{GeV}$) as \cite{CYf0K,CCYscalar}
\be
\mbox {Scenario I}: \bar f_{f_0(980)}&=&(0.37\pm0.02) \mbox {GeV}, \quad\bar f_{f_0(1500)}=-(0.255\pm0.03) \mbox {GeV},\non
B_1(980)&=&-0.78\pm0.08, \quad\quad\quad B_3(980)=0.02\pm0.07,\non
B_1(1500)&=&0.80\pm0.40, \quad\quad\;\quad B_3(1500)=-1.32\pm0.14;\non
\mbox{Scenario II}: \bar f_{f_0(1500)}&=&(0.49\pm0.05) \mbox {GeV},\non
B_1(1500)&=&-0.48\pm0.11,\quad\quad\quad B_3(1500)=-0.37\pm0.20.
\en

\begin{table}\caption{Input parameters used in the numerical calculation\cite{CCYscalar,pdg08}. }
\begin{center}
\begin{tabular}{c |cc}
\hline \hline
 Masses &$m_{f_0(980)}=0.98 \mbox{ GeV}$,   &$ m_0^{K}=1.7 \mbox{ GeV}$, \\ \
  & $m_{f_0(1500)}=1.5 \mbox{ GeV}$&$ M_{B_s} = 5.37 \mbox{ GeV}$,\\
 \hline
  Decay constants &$f_{B_s} = 0.23 \mbox{ GeV}$,  & $ f_{K} = 0.16
 \mbox{ GeV}$,\\
 \hline
Lifetimes &
$\tau_{B^0_s}=1.466\times 10^{-12}\mbox{ s}$,&\\
 \hline
$CKM$ &$V_{tb}=0.9997$, & $V_{td}=0.0082e^{-i21.6^{\circ}}$,\\
 &$V_{ud}=0.974$, & $V_{ub}=0.00367e^{-i60^{\circ}}$.\\
\hline \hline
\end{tabular}\label{para}
\end{center}
\end{table}

As for the explicit form of the Gegenbauer moments for the twist-3
distribution amplitudes $\Phi_{f_0}^S$ and $\Phi_{f_0}^T$, they have not
been studied in the literature, so we adopt the asymptotic form \be
\Phi^S_{f_0}&=& \frac{1}{2\sqrt {2N_c}}\bar f_{f_0},\,\,\,\,\,\,\,\Phi_{f_0}^T=
\frac{1}{2\sqrt {2N_c}}\bar f_{f_0}(1-2x). \en

In the each twist LCDAs, the appearance of Gegenbauer polynomials is from the expansion of
non-local operator into local conformal operators \cite{braun}. Sometimes, the twist-3 contributions are
important, because they can be enhanced due to some mechanism like chiral enhancement, especially in
the condition of the leading twist contributions being small or zero.

The twist-2 kaon distribution
amplitude $\Phi^{A}_{K}$, and the twist-3 ones $\Phi^{P}_{K}$ and
$\Phi^{T}_{K}$ have been parametrized as
\begin{eqnarray}
 \Phi_{K}^{A}(x) &=&  \frac{f_K}{2\sqrt{2N_c} }
    6x (1-x)
    \left[1+0.51(1-2x)+0.3(5(1-2x)^2-1)
  \right],\label{piw1}\\
 \Phi^{P}_{K}(x) &=&   \frac{f_K}{2\sqrt{2N_c} }
   \left[ 1+0.24(3(1-2x)^2-1)\right.\non &&\left.\qquad\qquad-0.12/8(3-30(1-2x)^2+35(1-2x)^4)\right]  ,\\
 \Phi^{T}_{K}(x) &=&  \frac{f_K}{2\sqrt{2N_c} } (1-2x)
   \left[ 1+0.35
   (1-10x+10x^2)\right] .\quad\quad\label{piw}
 \end{eqnarray}
The $B_s$ meson's wave function can be written as
\be
\phi_{B_s}(x,b)=N_{B_s}x^2(1-x)^2\exp[-\frac{M^2_{B_s}x^2}{2\omega^2_{b_s}}-\frac{1}{2}(\omega_{b_s}b)^2],
\en
where $\omega_{b_s}$ is a free parameter and we take $\omega_{b_s}=0.5\pm0.05$ GeV in numerical calculations,
and $N_{B_s}=63.67$ is the normalization factor for $\omega_{b_s}=0.5$ \cite{ali}.

For the numerical calculation, we list the other input parameters in Table
~I.

Using the wave functions and the values of relevant input parameters, we find the numerical values
of the corresponding form factors $\bar B^0_s\to f_0(s\bar s)$ at zero meomentum transfer
\be
F^{\bar B^0_s\to f_0(980)}_0(q^2=0)&=&0.33^{+0.02+0.02+0.02}_{-0.01-0.01-0.01},\quad\mbox{ scenario I},
\\ F^{\bar B^0_s\to f_0(1500)}_0(q^2=0)&=&-0.25^{+0.01+0.06+0.04}_{-0.00-0.05-0.03}, \quad\mbox{ scenario I},
\\F^{\bar B^0_s\to f_0(1500)}_0(q^2=0)&=&0.59^{+0.06+0.04+0.05}_{-0.06-0.03-0.05}, \quad\;\;\;\mbox{ scenario II},
\en
where the uncertainties are from the decay constant,
the Gegenbauer moments $B_1$ and $B_3$ of the meson $f_0$.
The large form factors result from the large decay constants of the scalar mesons. The opposite sign of the
$\bar B^0_s\to f_0(1500)(s\bar s)$ form factor in the
upper two scenarios arises from the decay constant of $f_0(1500)$. These values agree well with those as given in Ref.\cite{btos}

In the $B_s$-rest frame, the decay rate of $\bar B^0_s\to f_0K^0$ can be expressed as
\be
\Gamma=\frac{G_F^2}{32\pi m_{B_s}}|{\cal M}|^2(1-r^2_{f_0}),
\en
where $r_{f_0}=m_{f_0}/m_{B_s}$ and ${\cal M}$ is the decay amplitude of $B\to f_0K^0$, which has been given in equation (\ref{snamp}) .
If $f_0(980)$ and $f_0(1500)$ are purely composed of $n\bar n$($s\bar s$), the branching ratios
of $\bar B^0_s\to f_0(980)K^0, f_0(1500)K^0$ are
\be
{\cal B}(\bar B^0_s\to
f_0(980)(n\bar n)K^0)&=&(5.4^{+0.7+1.2+0.1}_{-0.5-1.0-0.0})\times 10^{-6}, \mbox{ scenario I},\\
{\cal B}(\bar B^0_s\to
f_0(1500)(n\bar n)K^0)&=&(4.4^{+0.1+3.4+0.5}_{-0.1-2.0-0.3})\times 10^{-6}, \mbox{ scenario I},\\
{\cal B}(\bar B^0_s\to
f_0(1500)(n\bar n)K^0)&=&(3.9^{+0.8+1.4+0.2}_{-0.7-1.3-0.1})\times 10^{-6}, \mbox{ scenario II};\\
{\cal B}(\bar B^0_s\to
f_0(980)(s\bar s)K^0)&=&(1.0^{+0.1+0.1+0.1}_{-0.1-0.2-0.1})\times 10^{-6}, \mbox{ scenario I},\\
{\cal B}(\bar B^0_s\to
f_0(1500)(s\bar s)K^0)&=&(0.2^{+0.00+0.30+0.06}_{-0.00-0.13-0.05})\times 10^{-6}, \mbox{ scenario I},\\
{\cal B}(\bar B^0_s\to
f_0(1500)(s\bar s)K^0)&=&(2.1^{+0.5+0.5+0.7}_{-0.4-0.4-0.5})\times 10^{-6}, \mbox{ scenario II},
\end{eqnarray}
where the uncertainties are from the same quantities as above.

\begin{figure}[t,b]
\begin{center}
\includegraphics[scale=0.7]{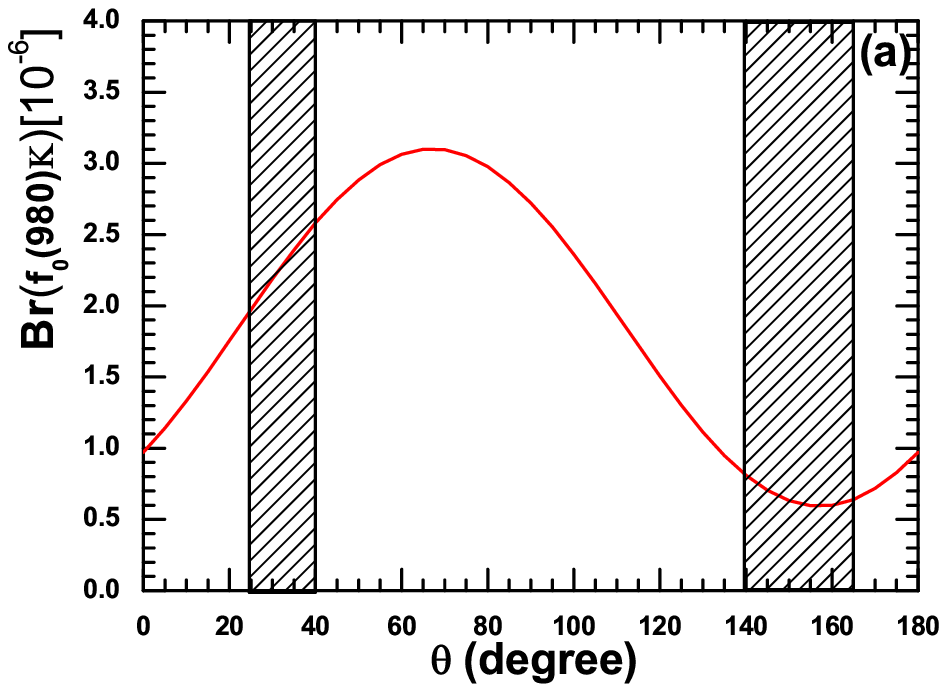}
\includegraphics[scale=0.7]{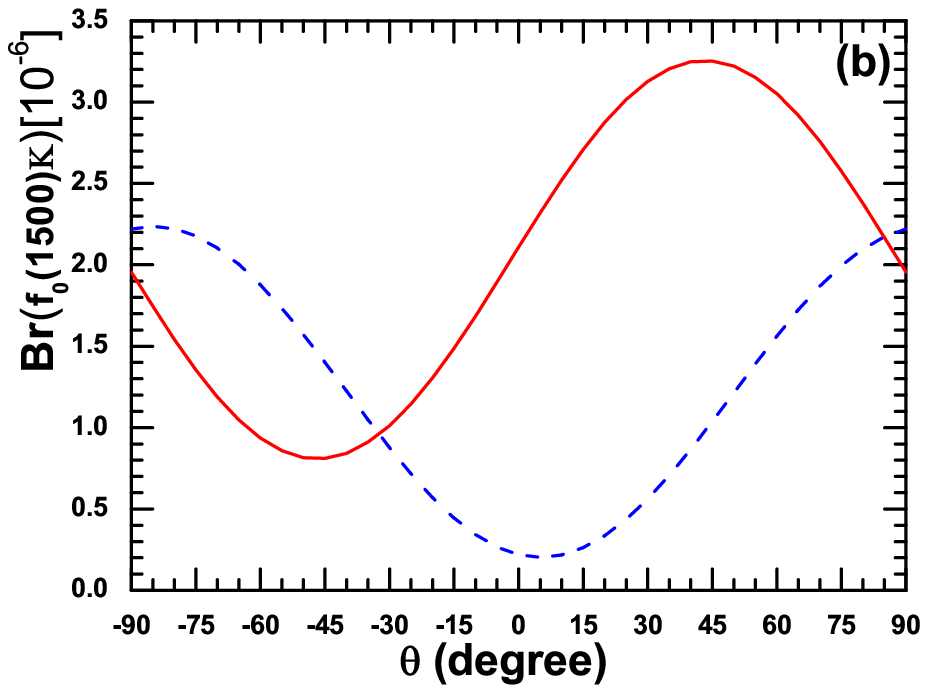}
\vspace{0.3cm} \caption{The dependence of the branching ratios for
$\bar B^0_s\to f_0(980)K^0$ (a) and $\bar B^0_s\to f_0(1500)K^0$ (b) on the
mixing angle $\theta$ using the inputs derived from QCD sum rules. For the right panel, the dashed (solid) curve is
plotted in scenario I (II).
The vertical bands show
two possible ranges of $\theta$: $25^\circ<\theta<40^\circ$ and $140^\circ<\theta<165^\circ$.}\label{fig2}
\end{center}
\end{figure}
\begin{table}
\caption{ Decay amplitudes for $\bar B^0_s\to f_0(980)K^0,f_0(1500)K^0$ ($\times 10^{-2} \mbox {GeV}^3$).}
\begin{center}
\begin{tabular}{cc|c|c|c|c|c|c}
\hline \hline  $\bar ss$&&$F^K_e$ & $M^K_e$ & $M^K_a$ & $F^K_a$ &$M^{f_0}_e$\\
\hline
$f_0(980)K^0$ (SI) &   &-1.0&$0.2+0.5i$&$-0.4-0.1i$&$2.2-10.3i$ &$-5.7-5.1i$\\
$f_0(1500)K^0$ (SI) &   &1.8&$-0.6+0.8i$&$-0.1-0.2i$&$6.6+13.4i$ &$-2.2-8.1i$\\
$f_0(1500)K^0$ (SII) &   &-2.4&$0.6+0.07i$&$-0.5-0.04i$&$-0.8-21.1i$ &$-6.3-1.7i$\\
\hline \hline
  $\bar nn $&&$F^{f_0}_e$&$M^{f_0,T}_e$ &  $M^{f_0}_e$ & $M^{f_0}_a$ & $F^{f_0}_a$  \\
\hline
$f_0(980)(\bar nn)K^0$ (SI)& &7.2&$57.5+47.6 i$& $-10.1-9.5i$&$0.4+0.4i$&$1.9-8.7i$\\
$f_0(1500)(\bar nn)K^0$ (SI)& &-7.6&$20.9+77.1 i$& $-2.6-15.1i$&$-0.7+0.7i$&$6.8+11.8i$\\
$f_0(1500)(\bar nn)K^0$ (SII)& &14.6&$57.9+14.3 i$& $-10.7-2.9i$&$0.9+0.1i$&$-1.4-17.8i$\\
\hline\hline
\end{tabular}\label{amp}
\end{center}
\end{table}

In table II, we list values of the factorizable and non-factorizable amplitudes from the emission
and annihilation topology diagrams of the decays $\bar B^0_s\to f_0(980)K, f_0(1500)K$. $F^K_{e(a)}$ and $M^K_{e(a)}$
are the $K$ emission (annihilation) factorizable
contributions and non-factorizable contributions from penguin operators respectively. Similarly,
$F^{f_0}_{e(a)}$ and $M^{f_0}_{e(a)}$ denote as
the contributions from $f_0$ emission (annihilation) factorizable contributions and non-factorizable
contributions from penguin operators respectively. It is easy to see that $F^{f_0}_{e(a)}$ and
$M^{f_0}_{e(a)}$ are larger than $F^K_{e(a)}$
and $M^K_{e(a)}$, that is $f_0$-
emission diagrams give large contributions. $M^{f_0,T}_{e}$ denotes the $f_0$ emission non-factorizable
contribution from tree operator $O_2$. From the table, one can find that
the contributions from $n\bar n$ component of $f_0$ are larger than those from $s\bar s$ component,
the one reason is that $M^{f_0,T}_{e}$ from the
operator $O_2$ is much larger than other amplitudes. Furthermore,
${\cal B}(\bar B^0_s\to f_0(n\bar n)K)$
is about two or five times of ${\cal B}(\bar B^0_s\to
f_0(s\bar s)K)$. Certainly, as to the decay $\bar B^0_s\to f_0(1500)K$ in SI, the difference
is greater. One can see the values of
${\cal B}(\bar B_s\to f_0(1500)(n\bar n)K)$ in SI and SII are close to each other,
but the values of ${\cal B}(\bar B^0_s\to f_0(1500)(s\bar s)K)$ in the two scenarios are much
differ, which results the total branching ratios
have an apparent difference between these two scenarios
(shown in figur~2(b)).

In figure~\ref{fig2}, we plot the branching ratios as functions of the
mixing angle $\theta$. If the mixing angle $\theta$ falls into the
range of $25^\circ<\theta<40^\circ$, the branching ratio of $\bar
B^0_s\to f_0(980)K$ is: \be 2.0\times 10^{-6}<{\cal B}(\bar B^0_s\to
f_0(980)K)<2.6\times 10^{-6}, \en while $\theta$ lies in the range
of $140^\circ<\theta<165^\circ$, ${\cal B}(\bar B^0_s\to f_0(980)K)$
is roughly $6.5\times 10^{-7}$. The dependence of the branch ratio of $\bar B^0_s\to f_0(980)K$ is strong in the whole mixing
angle range, but not sensitive in some ranges, for example, $140^\circ<\theta<165^\circ$. As to the decay $\bar
B^0_s\to f_0(1500)K$, because there are more discrepancies for the
structure of $f_0(1500)$, we do not show the possible allowed mixing angle range in
Fig.2(b). Lattice QCD predicted that the mass of the ground state
scalar glueball is around $1.5-1.8$ GeV \cite{glueball,cheny}, so the three mesons
$f_0(1370), f_0(1500)$ and $f_0(1710)$ become the potential
candidates. The mixing matrix can be written as \cite{wwang2}
\begin{eqnarray}
 \left( \begin{array}{c}
 f_0(1710) \\f_0(1500)\\ f_0(1370)
 \end{array} \right) = \left( \begin{array}{ccc} a_1 & a_2
&a_3 \\b_1 &b_2 &b_3  \\c_1 &c_2 &c_3\end{array}\right)
\left(\begin{array}{c} G \\ \bar ss \\ \bar nn\end{array}
\right).\label{mixing}
\end{eqnarray}
For each physical scalar meson, the corresponding component
coefficients satisfy the normalization condition, so we have $\sqrt
{|b_1|^2+|b_2|^2+|b_3|^2}=1$ for the meson $f_0(1500)$. For the
earlier lattice calculations predicting the scalar glueball mass to
be about $1550$ MeV and the decay width of $f_(1500)$ being not
compatible with a simple $q\bar q$ state \cite{amsler0}, Amsler and
Close claimed that $f_0(1500)$ is primarily a scalar glueball
\cite{amsler}. But in the $SU(3)$ symmetry limit, Cheng et al.
reanlyze all existing experimental data and find that $f_0(1500)$ is
a pure SU(3) octet with a very tiny glueball content. Their results
for the mixing coefficients are given in the following \cite{mixingch}
\begin{eqnarray}
 \left( \begin{array}{c}
 f_0(1710) \\f_0(1500)\\ f_0(1370)
 \end{array} \right) = \left( \begin{array}{ccc} 0.93 & 0.17
&0.32 \\-0.03 &0.84 &-0.54  \\-0.36 &0.52 &0.78\end{array}\right)
\left(\begin{array}{c} G \\ \bar ss \\ \bar nn\end{array}
\right).\label{mixing}
\end{eqnarray}
From above Equation, it is ease to see that $f_0(1710)$ is composed
primarily of a scalar glueball. This conclusion is also supported by
an improved LQCD calculation \cite{cheny}, which predicts
that the mass of the scalar glueball is $1710\pm50\pm80$ MeV. Here
we use the mixing coefficients for $f_0(1500)$ given in
Eq.(\ref{mixing}) and neglect the tiny component of glueball, that
is $\mid f_0(1500)\rangle=0.84\mid s\bar s
\rangle-0.54\mid n\bar n \rangle$. The branching ratios in two scenarios given as:
\be
{\cal
B}(\bar B^0_s\to f_0(1500)K^0)&=&9.7\times10^{-7}, \mbox{
Scenario I},\non {\cal B}(\bar B^0_s\to
f_0(1500)K^0)&=&9.5\times10^{-7}, \mbox{ Scenario II}, \en
which is obtained by the mixing angle taken as $-32.7^\circ$. From these results, it is
difficult to determine which scenario is more preferable than the other.

Now we turn to the evaluations of the CP-violating asymmetries of
the considered decays in PQCD approach.
For the neutral decays $\bar B^0_s\to f_0(980)K_S,f_0(1500)K_S$,
there are both direct $CP$ asymmetry $A^{dir}_{CP}$ and
mixing-induced $CP$ asymmetry $A^{mix}_{CP}$. The time dependent
$CP$ asymmetry of $B^0_s$ decay into a $CP$ eigenstate $f$ is defined
as
\be
{\cal A}_{CP}(t)={\cal A}^{dir}_{CP}(B^0_s\to f)\cos(\Delta m_s)+{\cal A}^{mix}_{CP}(B^0_s\to f)\sin(\Delta m_s),
\en
with
\be
{\cal A} ^{dir}_{CP}(B^0_s\to f)&=&\frac{|\lambda|^2-1}{1+|\lambda|^2},
\;\;\;
{\cal A}^{mix}_{CP}(B^0_s\to f)=\frac{2 Im \lambda}{1+|\lambda|^2},\\
\lambda&=&\eta e^{-2i\beta}\frac{{\cal A}(\bar B^0_s\to f)}{{\cal
A}(B^0_s\to f),}
\en
where  $\eta=\pm1$ depends on the $CP$ eigenvalue
of $f$, $\Delta m_s$ is the mass difference of the two neutral $B_s$
meson eigenstates. Here we only give the direct CP asymmetries.

\begin{table}
\caption{Direct $CP$ asymmetries (in units of \%) of $\bar B^0_s \to f_0(980)K^0, f_0(1500)K^0$ decays for $n\bar n$
and $s\bar s$ components, respectively.}\label{CP}
\begin{center}
\begin{tabular}{c|c|c|c}
   \hline \hline
   Channel & Scenario I & Scenario II &  \\
   \hline   $\bar B^0_s \to f_0(980)(n\bar n) K^0$&$-57.0$&-\\
    $\bar B^0_s \to f_0(980)(s\bar s) K^0$&$0$&-\\
 \hline     $\bar B^0_s \to f_0(1500)(n\bar n) K^0 $&$17.6$&$-94.8$\\
    $\bar B^0_s \to f_0(1500)(s\bar s)K^0 $&$0$&$0$\\
   \hline\hline
\end{tabular}
   \end{center}
\end{table}

\begin{figure}[t,b]
\begin{center}
\includegraphics[scale=0.7]{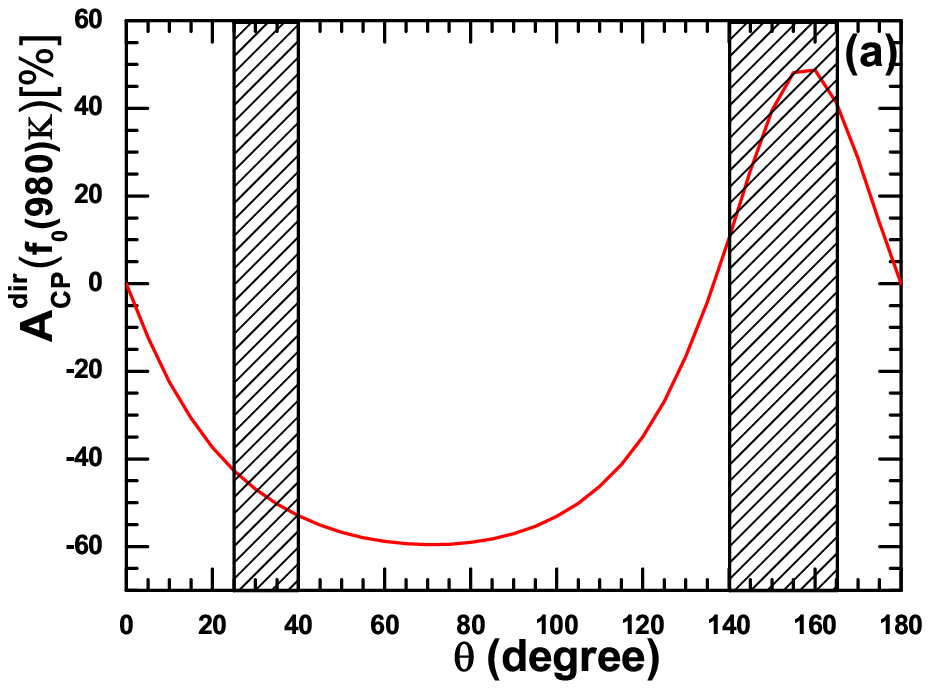}
\includegraphics[scale=0.7]{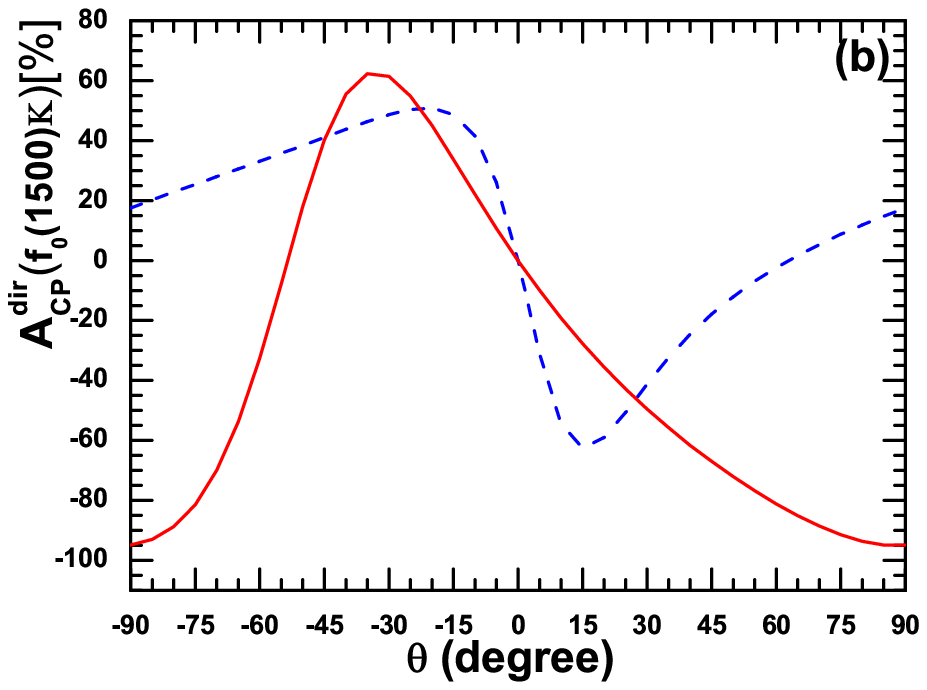}
\vspace{0.3cm} \caption{The dependence of the direct CP asymmetries for
$\bar B^0_s\to f_0(980)K$ (a) and $\bar B^0_s\to f_0(1500)K$ (b) on the
mixing angle $\theta$. For the right panel, the dashed (solid) curve is plotted in scenario I (II).
The vertical bands show two possible ranges of $\theta$: $25^\circ<\theta<40^\circ$ and $140^\circ<\theta<165^\circ$.}\label{fig3}
\end{center}
\end{figure}

In $\bar B^0_s\to f_0(s\bar s)K$, there is no tree contribution at the
leading order, so the direct CP asymmetry is naturally zero. As
$\bar B^0_s\to f_0(n\bar n)K$, the corresponding direct CP asymmetries
are listed in table III. For the decay $\bar B^0_s\to f_0(1500)K$, the
amplitudes from the non-factorizable $f_0$-emission and annihilation
topologies, that is $M_e^{f_0}$ and $F_a^{f_0}$, are constructive in
scenario II but are destructive in scenario I (seen in table II). Furthermore, the
contributions from the factorizable $f_0$-emission diagrams have an
opposite sign between scenario I and II (which is because that the decay
constant of $f_0(1500)$ is opposite in two scenarios). These reasons result that there exists a great difference for the direct
CP asymmetries of $\bar B^0_s\to f_0(1500)(n\bar n)K$
in two scenarios (shown in table III). The dependence of the direct
CP violating asymmetries for these decays are shown in Fig.3. From Fig.3(a), we can see that the signs of the direct CP
asymmetries in
the two allowed mixing angle ranges are opposite, it gives the hint that one can determine the mixing angle by comparing
with the future experimental results. For the decay $\bar B^0_s\to f_0(1500)K$, if we still
use the mixing scheme given by Eq.(\ref{mixing}) for $f_0(1500)$, the direct CP asymmetry is
\be
{\cal A} ^{dir}_{CP}(\bar B^0_s\to f_0(1500)K^0)&=&47.4\%, \qquad\mbox{
scenario I},\non
{\cal A} ^{dir}_{CP}(\bar B^0_s\to
f_0(1500)K^0)&=&62.8\%, \qquad \mbox{ scenario II}.
\en
Although the CP asymmetry for the decay $\bar B^0_s\to f_0(1500)K^0$ is large, it is difficult to measure it,
since its branching ratio is small.


\section{Conclusion}\label{summary}

In this paper, we calculate the branching ratios and the direct CP-violating
asymmetries of $ \bar B^0_s\to f_0(980)K, f_0(1500)K$ decays
in the PQCD factorization approach. From our calculations and phenomenological analysis, we find the following results:
\begin{itemize}
\item
In general, the contributions from $f_0$-emission diagrams are larger than those from $K$-emission diagrams. Especially,
the $f_0$-emission non-factorizable contribution from tree operator $O_2$ is quite larger than other amplitudes. For the decays
$\bar B^0_s\to f_0(980)K, f_0(1500)K$,
the contributions from $n\bar n$ component are larger than those from $s\bar s$ component in two scenarios.

\item
Using the wave functions and the values of relevant input parameters, we find the numerical results
of the corresponding form factors $\bar B^0_s\to f_0(s\bar s)$ at zero meomentum transfer
\be
F^{\bar B^0_s\to f_0(980)}_0(q^2=0)&=&0.33^{+0.02+0.02+0.02}_{-0.01-0.01-0.01},\quad\mbox{ scenario I},
\\ F^{\bar B^0_s\to f_0(1500)}_0(q^2=0)&=&-0.25^{+0.01+0.06+0.04}_{-0.00-0.05-0.03}, \quad\mbox{ scenario I},
\\F^{\bar B^0_s\to f_0(1500)}_0(q^2=0)&=&0.59^{+0.06+0.04+0.05}_{-0.06-0.03-0.05}, \quad\;\;\;\mbox{ scenario II}.
\en
The values of $F^{\bar B^0_s\to f_0(1500)}_0(q^2=0)$ for two scenarios can be used to identify which scenario is favored
by compare with the future experimental results.
\item
If the mixing angle $\theta$ falls into the range of $25^\circ<\theta<40^\circ$, the branching ratio of
$\bar B^0_s\to f_0(980)K$ is
\be
2.0\times 10^{-6}<{\cal B}(\bar B^0_s\to f_0(980)K)<2.6\times 10^{-6},
\en
while $\theta$ lies in the range of $140^\circ<\theta<165^\circ$, ${\cal B}(\bar B^0_s\to f_0(980)K)$ is about
$6.5\times 10^{-7}$.
\item
if we identify the meson $f_0(1500)$ as a pure SU(3) octet state and use the mixing scheme giving by $\mid f_0(1500)\rangle=0.84\mid s\bar s
\rangle-0.54\mid n\bar n \rangle$, one can find that the branching ratios in two scenarios are both close to $1.0\times10^{-6}$. Although the CP asymmetry
for the decay $\bar B^0_s\to f_0(1500)K^0$ is large, it is difficult to measure it, since its branching ratio is small.
\end{itemize}

\section*{Acknowledgment}
This work was supported by Foundation of Henan University of Technology under Grant No.150374. The author would like to thank
Hai-Yang Cheng, Cai-Dian L\"U, Wei Wang, Yu-Ming Wang for helpful discussions.

\end{document}